\documentclass[aps,twocolumn]{revtex4}

\usepackage{graphicx}
\usepackage{dcolumn}
\usepackage{bm}
\usepackage{amsmath}
\begin{document}

\newcommand{\bk}{{\bf k}}
\newcommand{\bp}{{\bf p}}
\newcommand{\bv}{{\bf v}}
\newcommand{\bq}{{\bf q}}
\newcommand{\tbq}{\tilde{\bf q}}
\newcommand{\tq}{\tilde{q}}
\newcommand{\bQ}{{\bf Q}}
\newcommand{\br}{{\bf r}}
\newcommand{\bR}{{\bf R}}
\newcommand{\bB}{{\bf B}}
\newcommand{\bA}{{\bf A}}
\newcommand{\bK}{{\bf K}}
\newcommand{\vd}{{v_\Delta}}
\newcommand{\tr}{{\rm Tr}}
\newcommand{\kslash}{\not\!k}
\newcommand{\qslash}{\not\!q}
\newcommand{\pslash}{\not\!p}
\newcommand{\rslash}{\not\!r}
\newcommand{\bs}{{\bar\sigma}}

\title{Crystalline electron pairs}

\author{M. Franz}
\affiliation{Department of Physics and Astronomy,
University of British Columbia, Vancouver, BC, Canada V6T 1Z1}
\date{\today}

\begin{abstract}
\end{abstract}
\maketitle

At low temperatures many compounds exhibit the phenomenon of superconductivity, a state of vanishing resistance to electrical current. In high-temperature cuprate superconductors the critical temperature $T_c$  can be as high as 160K, and the mechanism through which this obtains remains shrouded in mystery. A closely related fundamental question is the nature of the electronic state outside of the ``superconducting dome'' sketched in the figure. In the early days of high-$T_c$, many in the field harbored hopes that the region intermediate between the anti-ferromagnetic Mott insulator and the superconductor may contain exotic electron liquid phases with no broken symmetries and ``fractionalized'' elementary excitations \cite{anderson1}, akin to Luttinger liquids known to exist in one-dimensional interacting systems. Much of the subsequent activity in this field focused on searches for such exotic forms of electronic matter, but no convincing evidence has ever been found. 

An alternative to the fractionalized liquid is an ordered state. Recent high resolution scanning tunneling microscopy (STM) measurements of Hanaguri {\em et al.} \cite{han}, performed on a relative newcomer to the cuprate family, Ca$_{2-x}$Na$_x$CuO$_2$Cl$_2$  (Na-CCOC), offer an exciting glimpse of what lies to the left of the superconducting dome. The inset to the figure shows periodic pattern in the local electron density of states (LDOS) obtained on a single crystal of Na-CCOC with doping level close to $x_c=1/8$. Such pattern, whose period is independent of the energy of the tunneling electron, is strongly suggestive of underlying crystalline electronic order and comes on the heels of earlier experimental hints \cite{howald,vershinin}  that such static order may occur in another cuprate material Bi$_2$Sr$_2$CaCu$_2$O$_{8+\delta}$  (BiSCCO). 

There are essentially three ways to suppress superconductivity in a material, indicated by arrows in the figure. One can raise temperature above $T_c$, apply magnetic field $B$, or change the doping level $x$  by altering the chemical composition. Each method has its unique experimental challenges but based on the topology of the phase diagram one would expect to reach the same state of electronic matter upon exiting the superconducting state via any of the three methods, unless another phase boundary is encountered in the process. Moving along the doping axis, the results of Hanaguri {\em et al.} complete the triad of tests that convincingly demonstrate the existence of crystalline electronic order inside and outside of the superconducting dome. Historically first in this sequence, moving along the $B$  axis, was the discovery of checkerboard patterns in the LDOS in the vicinity of magnetic vortices by Hoffman {\em et al.} \cite{hoffman}. Next, exploring the $T$-direction, came the report of weak modulations deep inside the superconducting state by Howald {\em et al.} \cite{howald} followed by an even more convincing observation of similar patterns above $T_c$  by Vershinin {\em et al.} \cite{vershinin}. 

The new results of Hanaguri {\em et al.} are perhaps the most spectacular in this group in that the checkerboards completely dominate the STM signal and can be clearly seen in the raw data. This clarity, combined with atomic resolution, allows for detailed examination of the phenomenon. An intriguing feature of the data is that apparently the same phenomenon occurs in the insulating state (doping $x=0.08$ ) and in the superconducting state ($x=0.10$ and 0.12, with $T_c=15$K and 20K, respectively). This implies that superconductivity coexists with charge ordering in this material.

\begin{figure}
\includegraphics[width = 8cm]{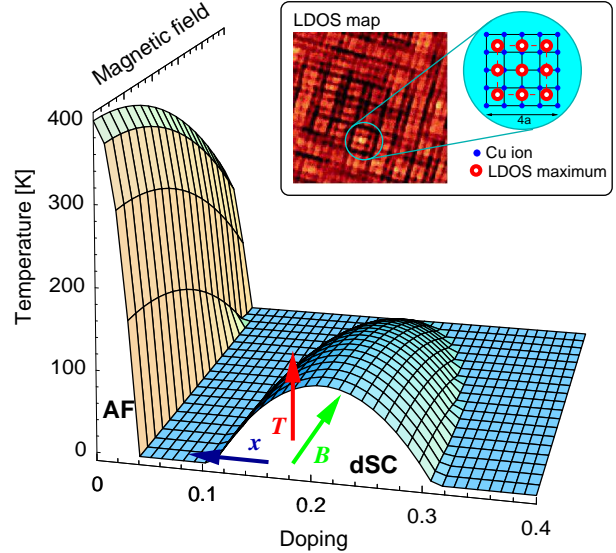}
\caption{Phase diagram of cuprates. AF denotes antiferromagnet and $d$SC a $d$-wave superconductor. Arrows indicate different ways to exit from the superconducting state that were utilized in probing the underlying normal state by STM \cite{han,howald,vershinin,hoffman}. Inset shows the experimental LDOS pattern of Hanaguri {\em et al.} \cite{han} which exhibits $4a\times 4a$  unit cell, where $a$ is the lattice constant. Each unit cell contains 9 maxima which, with the exception of the central one, are not registered to the Cu sites. The pattern, however, is commensurate with the underlying Cu lattice and the Fourier analysis indicates exact periodicity of $4a$. This is in contrast to the situation in BiSCCO, where the checkerboard is incommensurate with periodicity 4.3$-$4.7$a$ }
\end{figure}
What do we learn from this beautiful data? First, it would appear that the exotic fractionalized liquid states envisioned in the early theoretical works do not materialize in cuprates. Instead, more conventional ordered states of electronic matter are observed. The former conclusion has been anticipated for some time now \cite{bonn} but it was not clear which alternative ground state would be realized in cuprates. New insights provided by the STM clarify the situation considerably. Questions, however, abound.  Most prominently, one would like to understand what is the precise nature of the ordered state and what relationship, if any, it bears to the nearby superconducting state.

A brief reflection reveals that the observed order cannot be a simple charge density wave (CDW). The electron excitation spectra, also measured in STM, exhibit one universal feature: LDOS is always reduced near the Fermi level (the so called pseudogap behavior) with the minimum pinned to the Fermi energy $\epsilon_F$. An ordinary CDW produces a gap tied to the particular ordering wavevector {\bf Q} that is generically not pinned to $\epsilon_F$ over all of the Fermi surface. In fact, the shapes of the excitation spectra in the superconducting and insulating phases are essentially identical in Na-CCOC. This observation suggests that the two states are intimately related. A possible link is furnished by the idea, articulated early on by Emery and Kivelson \cite{emery}, that the pseudogap state may be understood as a phase-disordered superconductor. Superconducting order parameter $\Delta$   can be driven to zero by thermal or quantum fluctuations in its phase $\varphi$, while retaining  non-zero amplitude $|\Delta|$. This scenario is attractive as it automatically ensures that the pseudogap, being a direct descendant of the superconducting gap, remains pinned to the Fermi energy. Moreover, the spectral lineshapes are naturally very similar to those in the superconducting state, with sharp features washed out by fluctuations \cite{franz}.

Where do the checkerboards fit into this picture? According to the number-phase uncertainty principle \cite{tinkham}, which asserts that $\Delta\varphi\cdot\Delta N\geq 1$  (where $\Delta N$  and $\Delta\varphi$   represent the uncertainty in particle number and phase, respectively), phase fluctuations in a superconductor tend to suppress fluctuations in the local charge density. One way to accommodate such a reduction in charge fluctuations is to set up a periodic charge modulation, consisting of a wave in the Cooper pair density, dubbed ``pair density wave'' (PDW). An extreme form of this PDW is known as the Wigner crystal of Cooper pairs and has pairs localized in a lattice, much like ions in a solid. These interesting new forms of electronic matter were studied in recent theoretical works \cite{chen,tesanovic,anderson2} and found to capture some qualitative features of the experimental data. One intriguing consequence of the PDW hypothesis is a possibility of formation of a supersolid phase, conjectured previously to occur in solid $^4$He. Supersolid retains the crystalline order of the pair Wigner crystal but simultaneously exhibits superconductivity, presumably in this case due to the excess Cooper pairs which cannot be accommodated in the crystal. This picture could naturally explain another enduring mystery in cuprates, i.e. that the superfluid density is proportional to doping $x$  and not to the total electron density $1-x$.

Another intriguing theoretical proposal starts from the Mott insulator and envisions Wigner crystal of holes \cite{fu}. While both scenarios predict periodic checkerboard patterns, there are qualitative differences which will, in due time, allow for experimental validation of the correct picture.  

The author thanks J.C. Davis, A.P. Iyengar, T. Pereg-Barnea, Z. Tesanovic and 
A. Yazdani for helpful discussions.

\end{document}